# FullRecall: A Semantic Search-Based Ranking Approach for Maximizing Recall in Patent Retrieval

Amna Ali[1], Liyanage C. De Silva[1,2], Pg Emeroylariffion Abas[1, *]

[1]Faculty of Integrated Technolgoeis, Universiti Brunei Darussalam, Brunei Darussalam
[2]School of Digital Science, Universiti Brunei Darussalam, Brunei Darussalam

| ARTICLE INFO | ABSTRACT |
|---|---|
|  | Patent examiners and inventors face significant pressure to verify the originality and non-obviousness of inventions, and the intricate nature of patent data intensifies the challenges of patent retrieval. Therefore, there is a pressing need to devise cutting-edge retrieval strategies that can reliably achieve the desired recall. This study introduces FullRecall, a novel patent retrieval approach that effectively manages the complexity of patent data while maintaining the reliability of relevance matching and maximising recall. It leverages IPC-guided knowledge to generate informative phrases, which are processed to extract key information in the form of noun phrases characterising the query patent under observation. From these, the top k keyphrases are selected to construct a query for retrieving a focused subset of the dataset. This initial retrieval step achieves complete recall, successfully capturing all relevant documents. To further refine the results, a ranking scheme is applied to the retrieved subset, reducing its size while maintaining 100% recall. This multi-phase process demonstrates an effective strategy for balancing precision and recall in patent retrieval tasks. Comprehensive experiments were conducted, and the results were compared with baseline studies, namely HRR2 [1] and ReQ-ReC [2]. The proposed approach yielded superior results, achieving 100% recall in all five test cases. However, HRR2[1] recall values across the five test cases were 10%, 25%, 33.3%, 0%, and 14.29%, while ReQ-ReC [2] showed 50% for the first test case, 25% for the second test case, and 0% for the third, fourth, and fifth test cases. The 100% recall ensures that no relevant prior art is overlooked, thereby strengthening the patent pre-filing and examination processes, hence reducing potential legal risks. |

## 1. Introduction

The requirement for comprehensive and meticulous search scans is significantly higher in patent information retrieval systems compared to traditional information retrieval systems [3], [4]. In traditional settings, retrieval systems often emphasize precision; delivering the most relevant results whilst allowing the exclusion of some relevant results [1]. However, in high-risk domains such as patent retrieval, where the consequences of disregarding important information are exorbitant, this trade-off is unacceptable. Missing even a single relevant document can result in consequences, such as invalid claims, rejected applications, litigation costs, or reputational damage [5], [6], [7], [8]. Consequently, recall, the ability to retrieve all relevant documents, is not just preferred, but also critical in the patent domain [9], [10].

Patent retrieval is thus inherently recall-oriented. For inventors, a high-recall search helps in assessing patentability, avoiding unnecessary filing expenses, and strengthening claims, whilst for patent examiners, it ensures accurate determination of novelty and inventive step. The growing volume and complexity of patent applications, however, make high-recall retrieval a technically challenging task [11]. Furthermore, the language in patents is often highly technical, domain-specific, and legally structured [12]; hence, difficult to understand and interpret. This complexity complicates both the retrieval process and query construction.

To support inventors in determining patentability and patent examiners in determining the validity of a patent application, comprehensive prior art searches are essential. These searches aim to capture relevant documents that could affect the novelty or inventiveness of the application. However, patents often contain additional and complex data, making it difficult to extract only the most relevant information. To overcome this, researchers have focused on enhancing both the search process and the formulation of the query itself [13], [14], [15], [16]. Modern methods attempt to distill essential information from the patent text to create more effective representations, but this remains a challenge due to the legalistic and domain-specific language often used in patents [12].

To improve retrieval outcomes, researchers have explored various methods including query expansion [17],[18], query formulation [19], relevance feedback [20],[21], and document clustering [22],[21]. Over time, these have evolved into more advanced methods leveraging machine learning (ML) and natural language processing (NLP), especially contextual models such as Bidirectional Encoder Representations from Transformers (BERT) and Generative Pre-trained Transformer (GPT), which can better comprehend the meaning of search requests [23], [1], [24],[25], [26].

Nevertheless, many keyword-based methods still suffer from semantic mismatches, where vocabulary differences between the query and relevant documents prevent retrieval [5], [27]. Reference [27], for instance, highlights how terminological inconsistencies and linguistic variability pose persistent barriers to effective patent retrieval. To address these limitations, modern techniques incorporate syntactic and semantic analysis [28], [29], [30], [31]. The authors in reference [28] propose a weakly-supervised deep neural network that deploys a full NLP pipeline along with a domain-specific BERT model tailored for patents. They demonstrate that this method produces semantically meaningful keywords, enhancing prior art retrieval effectiveness. Reference [32] demonstrates that retrieval quality can be enhanced through query formulation strategies that adapt to patent-specific structures. Meanwhile, reference [33] evaluates multiple keyword extraction methods, such as Term Frequency–Inverse Document Frequency (TF-IDF), TextRank, KeyBERT, and GPT-3.5; concluding that large language models like GPT-3.5 yield the most meaningful keywords. Similarly, reference [34] demonstrates that incorporating syntactic structures in keyword extraction improves results over conventional TF-IDF methods. Despite these advancements, challenges remain. Embedding-based retrieval using transformer models such as BERT [29], Sentence-BERT (SBERT) [31], and domain-specific models like BERT for Patents [26] provides deeper semantic understanding but requires significant computational resources and large labeled datasets [35], [5]. Recent studies have also proposed various re-ranking techniques. Neural re-ranking [36], BERT combined with BM25 [37], SBERT integrated with BM25 [38], graph-based scoring [39], and neural meta-embedding approaches [40] aim to boost relevance while maintaining or improving recall. Cross-lingual retrieval enhancements and ranking fusion methods have also been tested. However, balancing computational feasibility with high recall remains difficult [5], [41]. Another recent study [42] proposes SEARCHFORMER, a patent-specific Sentence-BERT model fine-tuned on real-world prior art citations. The authors claim that their proposed approach demonstrates statistically significant improvements over BM25 and other generic language models, confirming its effectiveness for semantic retrieval in patents. However, they utilized a limited evaluation set instead of full-corpus searches due to computational constraints. Additionally, this study does not report recall-based results of the patent retrievals. In yet another recent study [43], authors proposed a novel model, PK-Bert, that integrates a patent knowledge graph into a BERT-based Transformer framework. The authors claim that their proposed model enhances citation recommendations for patent examiners. Although the proposed model outperforms both standard BERT and K-Bert in citation prediction tasks, however, it fails to demonstrate the high recall required to retrieve maximum relevant citations.

Given the limitations of existing methods, this study proposes a novel blended patent retrieval approach, FullRecall, designed to maximize recall while maintaining relevance and interpretability. The proposed FullRecall framework is structured in three sequential phases to ensure the captures of relevant prior art. In the first phase, the framework performs a semantic similarity search to extract and rank key phrases from the query patent; with the key phrases selected based on prior knowledge from the description of the International Patent Classification (IPC) codes. This step condenses the complex language of the query patent into a set of prioritized semantic units. The second phase carries out a targeted keyword-based search using the ranked key phrases from the first phase, to extract a narrowed subset of relevant documents for further analysis. In the final phase, another semantic similarity search is carried out to refine the identified relevant documents, before deploying a ranking scheme to generate the final set of matched documents.

The main contribution of this study is the development of FullRecall framework, an NLP-based blended patent retrieval framework that integrates semantic similarity search, keyword-based filtering, and relevance-based re-ranking to facilitate comprehensive prior art searches. The proposed FullRecall framework is designed to support both inventors and patent examiners by addressing the limitations of existing approaches and ensuring no relevant prior art is overlooked. Experimental results show that FullRecall consistently achieves 100 percent recall across tested scenarios and outperforms benchmark methods such as HRR2 [1] and ReQ-ReC [2], while also maintaining meaningful document ranking to support efficient review.

The following are the main objectives of this study:

- Exploit natural language processing techniques to develop a high-recall framework for prior art retrieval to support inventors in assessing the novelty of their inventions during the pre-patentability phase, and to assist patent examiners in conducting comprehensive prior art searches for evaluating novelty and inventive step.
- Multistep retrieval and refinement of the patent data to maximise recall, simultaneously balancing precision by integrating semantic similarity, keyword-based filtering, and relevance-based re-ranking techniques into a unified retrieval pipeline.
- Evaluate the effectiveness of the proposed framework in terms of recall and ranking quality.
- Compare the performance of the proposed framework with existing state-of-the-art retrieval models

The remainder of this paper is organized as follows. Section 2 describes the FullRecall framework in detail, outlining its three-phase architecture and underlying techniques. Section 3 presents the experimental setup along with an in-depth analysis of the results. Section 4 provides a comparative evaluation between FullRecall and existing baseline approaches, namely HRR2 [1] and ReQ-ReC [2]. Finally, Section 5 concludes the paper with a summary of key findings and suggestions for future research.

## 2. Methodology: The FullRecall Patent Retrieval Framework

The FullRecall framework aims to improve patent prior art retrieval by ensuring comprehensive recall without compromising on relevance. As illustrated in Figure 1, the framework comprises three (3) sequential phases:

1. Feature extraction - ranked noun phrases

2. Intermediary intervention - conveyor information flow
3. Full recall - ranked relevant documents

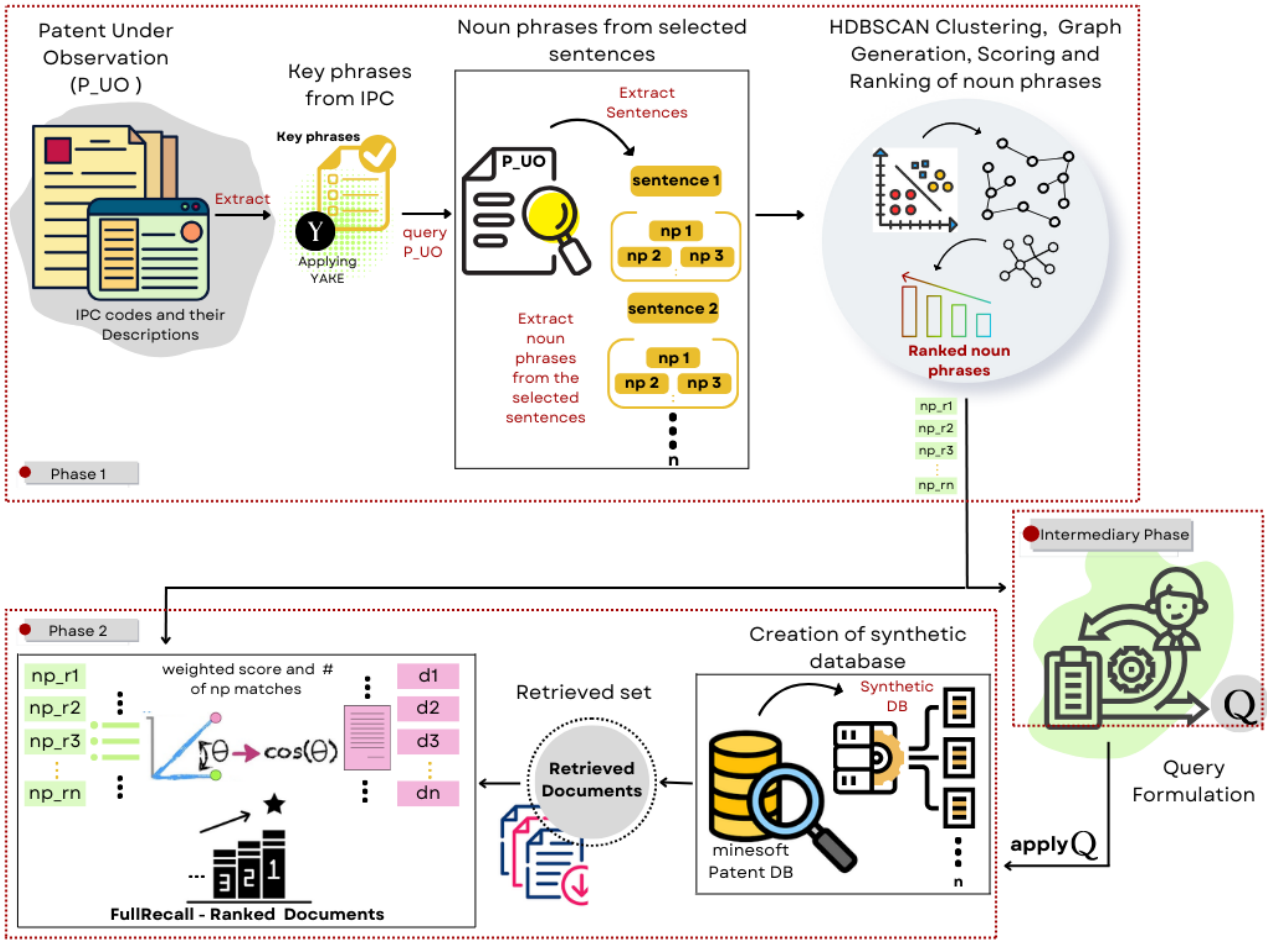

**Fig. 1**: Step-by-step overview of the proposed scheme

The process begins with a patent under observation that requires a comprehensive prior art search, either by an inventor assessing patentability before submission or an examiner evaluating an application post-filing. Patentability is evaluated based on novelty, inventive step, and industrial applicability; all of which necessitate a thorough review of existing prior art.

To support this process, the framework first identifies relevant IPC codes associated with the patent under observation. These codes provide structured descriptions that serve as prior knowledge for extracting domain-specific key phrases. A semantic similarity search is then conducted on the query patent text to locate key sentences that encapsulate its fundamental assertions. These selected sentences serve as the basis for noun phrase extraction, isolating core ingredient of technicality within these sentences. The extracted noun phrases are then clustered using Hierarchical Density-Based Spatial Clustering of Applications with Noise (HDBSCAN), an unsupervised density-based clustering algorithm, and ranked through a graph-based scoring scheme. This prioritization identifies the most informative and contextually relevant terms for query formulation.

In the intermediary phase, human oversight is introduced to refine the search process. The ranked noun phrases are used in constructing a refined search query. This intermediary step leverages domain expertise to ensure the query reflects the core inventive concepts, and thereby, enhancing the precision of retrieval.

In the third phase, a dataset is designed to emulate the actual database context, containing documents relevant to the patent application under observation. These documents are selected based on the IPC codes associated with the patent application under observation, ensuring domain consistency. The refined search query formulated in phase two is then executed against this dataset, to retrieve a refined set of patent documents that includes all known relevant prior art. A weighted scoring mechanism is applied to evaluate the semantic similarity between each retrieved document and the ranked noun phrases generated in the first phase. The documents are subsequently ranked in descending order of relevance, thereby ensuring a high recall rate while maintaining a concise and meaningful result set.

Figure 1 illustrates the FullRecall process, demonstrating how each phase builds upon the previous to achieve high-recall, high-relevance prior art retrieval.

*2.1. Feature extraction - Ranked noun phrases*

A patent under observation $P_{UO}$ requires a comprehensive prior art search to identify relevant documents, denoted as $Retrieved\ ranked_{patents}$. Whether for an inventor assessing patentability prior to filing, or for an examiner evaluating an application, this process involves navigating vast patent repositories. The FullRecall framework initiates this process by extracting key technical phrases from structured descriptions associated with IPC codes, which categorize patents based on technological domains; thereby, offering a focused search space for prior art retrieval whilst reducing irrelevant results [44].

The patent $P_{UO}$ is associated with a set of n IPC codes:

$$IPC_{P_{uo}} = \{IPC_1, IPC_2, \ldots, IPC_n\} \tag{1}$$

with each $IPC_i$ represents an IPC group consisting of $m_i$ subgroups:

$$S_{IPC,i} = \{S_{IPC_{i1}}, S_{IPC_{i2}}, \ldots S_{IPC_{im_i}}\} \tag{2}$$

where $S_{IPC_{ij}}$ represents the jth subgroup of group $IPC_i$. Each IPC group and subgroup has associated descriptions:

$D_{IPC_i}$: Description associated with classification group $IPC_i$.

$D_{S\_IPC_{ij}}$: Description associated with subgroup $S_{IPC_{ij}}$.

Given that $D_i$ represents description from $IPC_i$ and its subgroup, all descriptions associated with the patent $P_{UO}$ can be aggregated into:

$$D = \bigcup_{i=1}^{n} D_i \tag{3}$$

Where $D_i = D_{IPC_i} \cup \bigcup_{j=1}^{m_i} \{D_{S_{IPC_{ij}}}\}$. To extract meaningful key phrases, Yet Another Keyword Extractor (YAKE) is used to generate bi- and tri-gram key phrases ($kp_1$ and $kp_2$, respectively) from each description $D_i \in D$.

$$Y(D_i) = \{kp_1, kp_2\} \tag{4}$$

The set of ranked key phrases $K_{phrases}$ and set of unique ranked key phrases $K_{u\_phrases}$ after removal of duplicate key phrases are given by:

$$K_{phrases} = \bigcup_{D_i \in D} Y(D_i) \tag{5}$$

$$K_{u\_phrases} = \{x | x \in K_{phrases}\} \tag{6}$$

Key phrases in set $K_{u\_phrases}$ then serve as prior knowledge for the patent under observation $P_{UO}$.

Given the dense and technical nature of patent descriptions, it is essential to identify the most semantically informative sentences within $P_{UO}$. Let the set of all sentences in $P_{UO}$ be represented by:

$$S_{P_{UO}} = \{s_1, s_2, \ldots, s_k\} \tag{7}$$

To isolate the sentences that best reflect the core technical content, a semantic similarity scoring mechanism is applied. Specifically, each sentence $s_j \in S_{p_{UO}}$ is compared to a key phrase $kp_i$, using cosine similarity:

$$cos(kp_i, s_j) = \frac{kp_i \cdot s_j}{\|kp_i\| \|s_j\|} \quad \forall s_j \in S_{p_{UO}} \tag{8}$$

A sentence $s_j \in S_{p_{UO}}$ is retained in the refined sentence set $S_{selected}$ if its cosine similarity score with any key phrase exceeds a predefined threshold:

$$S_{selected} = \{s_j \in S_{P_{UO}} | cos(kp_i, s_j) > threshold\} \tag{9}$$

As more than 90% of the technical phrases used in patent documents are noun-based [44], noun phrase extraction is performed on $S_{selected}$, resulting in a set of ranked noun phrases:

$$N_{phrases} = \{np_1, np_2, \ldots, np_t\} \tag{10}$$

After removing duplicates noun phrases, the unique set $N_{u_{phrases}}$ of noun phrases is obtained:

$$N_{u_{phrases}} = \{x | x \in N_{phrases}\} \tag{11}$$

Each noun phrase is then embedded into a high-dimensional vector space to facilitate clustering and semantic comparisons:

$$Em_{u_{phrases}} = \{em_1, em_2, \ldots, em_k\} \tag{12}$$

where $em_i \in \mathbb{R}^d$ and $d = 384$, representing the dimensionality of the embedding space. Once the noun phrases have been embedded into a vector space, HDBSCAN is then used to group semantically similar noun phrases, resulting in a set $C$ of m distinct clusters:

$$C = \{c_1, c_2, \ldots, c_m\} \tag{13}$$

These clusters are non-overlapping and collectively cover the set of unique noun phrases:

$$\bigcup_{j=1}^{m} c_j = N_{u_{phrases}} \quad \text{with } c_i \cap c_j = \phi \text{ for } i \neq j \tag{14}$$

Next, a similarity graph $G = (V, E)$ is constructed. Each node $v_i \in V$ corresponds to a noun phrase $np_i \in N_{u_{phrases}}$ and an edge $e_{ij} \in E$ connects two nodes $v_i$ and $v_j$ if their corresponding embeddings exhibit cosine similarity above a specified threshold $\tau$:

$$\cos(em_i, em_j) > \tau \tag{15}$$

For each cluster $c_j$, a cluster centroid $em_{c_j}$ is computed as the mean (average) of all phrase embeddings within that cluster:

$$em_{c_j} = \frac{1}{|c_j|} \sum_{em_i \in c_j} em_i \tag{16}$$

where $|c_j|$ represents the number of noun phrase embeddings in cluster $c_j$. This centroid captures the central semantic position of the cluster in the embedding space. To assess the representativeness of each noun phrase in relation to the entire phrase space, a global graph centroid is computed by averaging the embeddings of all noun phrases in the unique set $N_{u_{phrases}}$:

$$em_{graph} = \frac{1}{|N_{u_{phrases}}|} \sum_{em_i \in N_{u_{phrases}}} em_i \tag{17}$$

where $|N_{u_{phrases}}|$ represents the number of noun phrase in the unique set $N_{u_{phrases}}$. These centroids, both local (cluster-level) and global, serve as reference points to evaluate the centrality and distinctiveness of each noun phrase. Ranking is then performed to prioritize noun phrases that are both representative and semantically informative. This is achieved by integrating structural and semantic indicators, with the computation of two scores for each noun phrase $np_i \in N_{u_{phrases}}$: the connectivity score $C_{np_i}$ and the uniqueness score $U_{np_i}$, as adapted from reference [44].

The connectivity score quantifies how strongly a noun phrase is embedded within its cluster and the overall graph structure; with a phrase that is closer to its cluster centroid and graph centroid considered more representative of the dataset. Connectivity score $C_{np_i}$ for noun phrase $np_i \in N_{u_{phrases}}$ is computed as:

$$C_{np_i} = \alpha_c \cdot Cc_{np_i} + (1 - \alpha_c) \cdot Cg_{np_i} \tag{18}$$

where $Cc_{np_i}$ is the Euclidean distance between the embedding $em_i$ of noun phrase $np_i \in c_j$ and the centroid $em_{c_j}$ of its cluster:

$$Cc_{np_i} = \|em_i - em_{c_j}\| \tag{19}$$

and $Cg_{np_i}$ is the Euclidean distance to the global centroid $em_{graph}$:

$$Cg_{np_i} = d_{graph}(np_i) = \|em_i - em_{graph}\| \tag{20}$$

Lower values of $Cc_{np_i}$ and $Cg_{np_i}$ indicate that a noun phrase is more central and thus more likely to be conceptually representative of both its immediate context and the entire dataset. The connectivity parameter $\alpha_c$ controls the balance between these two measures, allowing for fine-tuning of how much weight is given to local cluster coherence versus global dataset representation. A higher connectivity score indicates that a noun phrase is central to its cluster while also being well-integrated within the entire graph, making it a strong candidate for retrieval.

However, centrality alone does not ensure uniqueness. To address potential redundancy, a uniqueness score $U_{np_i}$ is computed to measure the distinctiveness of each noun phrase $np_i \in c_j$ within its own cluster $c_j$ and relative to the overall graph structure. This score is computed as:

$$U_{np_i} = \alpha_u \cdot Uc_{np_i} + (1 - \alpha_u) \cdot Ug_{np_i} \tag{21}$$

where $Uc_{np_i}$ is the minimum cosine similarity between the noun phrase $np_i \in c_j$ and other phrases within the same cluster $c_j$:

$$Uc_{np_i} = \min_{np_k \in c_j, np_k \neq np_i} \cos(em_i, em_k), \tag{22}$$

$Ug_{np_i}$ is the minimum similarity between the noun phrase $np_i \in c_j$ and the centroids of all other clusters, capturing cross-cluster uniqueness:

$$Ug_{np_i} = \min_{np_i \notin c_k} \cos(em_i, em_{c_k}) \tag{23}$$

and the uniqueness parameter $\alpha_u$ adjusts the emphasis on intra-cluster uniqueness versus cross-cluster distinctiveness, striking a balance between eliminating redundancy and preserving meaningful terms. A lower similarity score reflects greater semantic uniqueness, suggesting that the phrase conveys unique information distinct from both its immediate neighbors and broader conceptual clusters.

With both semantic aspects quantified, an overall semantic score $S(np_i)$ for noun phrase $np_i \in N_{u_{phrases}}$ is computed as the average of the connectivity $C_{np_i}$ and uniqueness $U_{np_i}$ scores:

$$S(np_i) = \frac{C_{np_i} + U_{np_i}}{2} \tag{24}$$

This semantic score provides a balanced measure of how central and distinctive a noun phrase is within the dataset and lays the foundation for more advanced structural scoring.

To enhance retrieval relevance, the semantic score is then supplemented with graph-theoretic metrics derived from the phrase similarity graph G. Specifically, three centrality measures: PageRank, Degree Centrality, and Betweenness Centrality, are employed to evaluate the structural importance of each noun phrase. These measures capture different aspects of a phrase's role within the graph, including its influence, connectivity, and ability to bridge distinct clusters of technical concepts.

PageRank score $P(np_i)$ captures the influence of each phrase within the graph, based on the likelihood of reaching that phrase through a random walk across the network. For a noun phrase $np_i \in N_{u_{phrases}}$, the score is computed by considering the PageRank scores of its directly connected neighbors $np_k \in \mathcal{N}(np_i)$:

$$P(np_i) = \sum_{np_k \in \mathcal{N}(np_i)} \frac{P(np_k)}{|E_{np_k}|} \tag{25}$$

where $\mathcal{N}(np_i)$ represents the set of noun phrases directly connected to $np_i$ in the graph G, and $|E_{np_k}|$ is the total number of edges (connections) from node $np_k$. A phrase that is highly connected to other influential phrases will receive a higher PageRank score, reflecting its central role in the dataset. While PageRank accounts for global influence within the dataset, Degree Centrality focuses on direct connectivity; with the score measuring the number of direct connections a noun phrase has with other phrases in the graph. Degree centrality $D(np_i)$ of a noun phrase $np_i \in N_{u_{phrases}}$ is computed as the normalized degree of a node, defined by the number of direct connections:

$$D(np_i) = \frac{|\mathcal{N}(np_i)|}{\max_{np_k \in N_{u_{phrases}}} |\mathcal{N}(np_k)|} \tag{26}$$

Normalization ensures comparability across datasets by scaling scores to the [0,1] range; allowing for consistent ranking across datasets of different sizes. A higher degree centrality indicates that a phrase appears frequently in different technological areas, making it a commonly referenced term.

Betweenness Centrality $B(np_i)$ quantifies how often a noun phrase acts as a bridge between different concepts; with a phrase with high Betweenness Centrality frequently appearing on shortest paths between other phrases, and thus, frequently serving as a semantic bridge linking distinct clusters of technical knowledge. For a noun phrase $np_i \in N_{u_{phrases}}$, Betweenness centrality $B(np_i)$ is calculated as:

$$B(np_i) = \sum_{s,t \in N_{u_{phrases}}, s \neq t} \frac{\sigma(s,t|np_i)}{\sigma(s,t)} \tag{27}$$

where $\sigma(s,t)$ represents the total number of shortest paths between phrases s and t, with $\sigma(s,t|np_i)$ representing the number of such paths passing through noun phrase $np_i$ only.

Finally, these structural metrics are integrated with the semantic score to produce a composite final ranking $R(np_i)$ for each noun phrase $np_i \in N_{u_{phrases}}$:

$$R(np_i) = \alpha.P(np_i) + \beta.D(np_i) + \delta.B(np_i) + (1 - \alpha - \beta - \delta) S(np_i) \tag{28}$$

where the parameters $\alpha$, $\beta$, and $\delta$ control the relative influence of PageRank, Degree Centrality, and Betweenness Centrality, while ensuring that the semantic score also contributes meaningfully to the final ranking. This weighted ranking ensures that both semantic distinctiveness and structural prominence are considered, enabling the selection of the most informative and representative terms for query formulation and prior art retrieval. The ranked noun phrases obtained from this process serve as the foundation for query formulation and document retrieval in the next phase, helping to refine the prior art search and improve retrieval accuracy.

## 2.2. Intermediary intervention - Conveyor information flow

In this phase, a refined structured query is formulated using the top K noun phrases from the ranked list, ensuring that only the most important and representative technical terms are included. These phrases are selected based on their final ranking scores, capturing both semantic significance and structural importance.

To maintain a balance between specificity and coverage, the number of selected phrases is constrained such that only the top-K highest-ranked noun phrases from the set of all extracted noun phrases $np_i \in N_{u_{phrases}}$, where $12 \leq K \geq 16$. This range ensures a manageable and effective query length while preserving retrieval quality. The resulting set of top-ranked phrases is defined as:

$$NP_{query} = \{np_i | np_i \in Top_K(R(N_{u_{phrases}})), 12 \leq K \geq 16\} \tag{29}$$

where $R(N_{u_{phrases}})$ represents the final ranking scores of the complete set of noun phrases $N_{u_{phrases}}$, with the function $Top_K(.)$ selecting only the K highest-ranked noun phrases. Once selected, the noun phrases in $NP_{query}$ are manually structured into a coherent search query:

$$Q_k = f(NP_{query}) \tag{30}$$

where the function $f(NP_{query})$ denotes the transformation function that formats the selected noun phrases into a structured search string. This step involves human oversight to ensure the query reflects the contextual nuances of the original patent document while remaining optimized for retrieving relevant prior art. The intermediary intervention thus acts as a conveyor of expert judgment, improving the accuracy and completeness of the subsequent retrieval process.

*2.3. Full recall - Ranked relevant documents*

The final phase ensures comprehensive retrieval of relevant prior art by ranking patent documents based on their semantic similarity to the query patent. This retrieval process differs slightly depending on whether it is performed by a patent examiner or an inventor, reflecting differences in the availability of IPC codes at the time of query formulation.

For patent examiners, IPC codes associated with the patent under observation $P_{UO}$ are already assigned. These classifications guide the construction of a candidate dataset $S_d$, consisting of patent documents sharing the same IPC codes as the query patent. The structured query $Q_k$, formulated in the previous phase is then applied to this dataset $S_d$ to retrieve a set of potentially relevant documents:

$$R_{sd} = Search(S_d, Q_k) \quad (31)$$

For inventors, the scenario somewhat differs, as a newly proposed invention has yet been assigned predefined IPC codes. In such cases, the inventor must first determine the most appropriate IPC classes. This can be achieved through:

- Reviewing IPC codes of similar patents; or
- Using AI-based IPC classification tools, such as WIPO IPCCAT or machine-learning-based tools such as BERT for Patents.

Once appropriate IPC codes have been identified, inventors can proceed with the same document retrieval process used by examiners.

Following retrieval, each document within the retrieved set $R_{sd}$ is analyzed with a focus on its independent claims, which play crucial roles in patent examinations as they define the core inventive concepts without referencing other claims. These claims also provide the most general and broadest protection sought for the invention [45]. Figure 2 summarizes the significance of independent claims of a patent document.

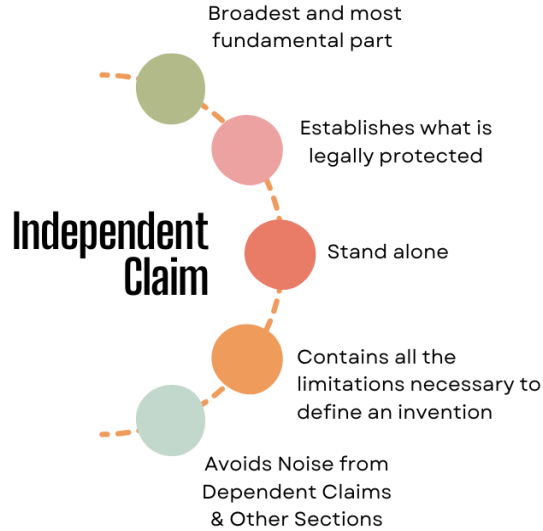

**Figure 2:** Significance of Independent Claims of a Patent Document

For each retrieved patent document $d_p \in R_{sd}$ and for each noun phrase $np_i \in NP_{query}$ from the ranked set of extracted noun phrases, the cosine similarity is computed between their respective embeddings:

$$\cos\left(e_{d_p}, e_{np_i}\right) = \frac{e_{d_p} \cdot e_{np_i}}{\left\|e_{d_p}\right\| \cdot \left\|e_{np_i}\right\|} \quad (32)$$

where $e_{d_p}$ represents the embedding of the retrieved patent document $d_p \in R_{sd}$, while $e_{np_i}$ represents the embedding of the noun phrase $np_i \in NP_{query}$; both derived using BERT for Patents with each embedding having a size of 768 dimensional vectors. To filter out weak matches, only document-phrase pairs $(e_{d_p}, e_{np_i})$ with cosine similarity exceeding a predefined threshold $\tau$ are considered:

$$valid\left(e_{d_p}, e_{np_i}\right) = \begin{cases} 1, & \text{if } \cos\left(e_{d_p}, e_{np_i}\right) > \tau \\ 0, & \text{otherwise} \end{cases} \quad (33)$$

The number of valid matches $NP_{matches}(d_p)$ for each document $d_p \in R_{sd}$ is then computed as:

$$NP_{matches}(d_p) = \sum_k valid(e_{d_p}, e_{np_k}) \quad (34)$$

To prioritize documents with higher semantic alignment, the valid matches are first sorted in descending order of cosine similarity. A weighted score for a document $d_p \in R_{sd}$ is computed to reward higher-ranked matches more heavily, with diminishing contribution from lower-ranked matches, using a $\frac{1}{k}$ penalty:

$$weighted_{score}(d_p) = \sum_{k=1}^{NP_{matches}(d_p)} \frac{\cos\left(e_{d_p}, e_{np_i}^{(k)}\right)}{k} \quad (35)$$

To further enhance ranking fidelity, the final document score integrates both the weighted score and the total number of valid matches, moderated by a tunable weight parameter $\lambda$:

$$final_{score}(d_p) = weighted_{score}(d_p) + \lambda \cdot NP_{matches}(d_p) \quad (36)$$

Finally, all retrieved documents are sorted in descending order based on their final scores to prioritize those that exhibit the strongest semantic match with the query patent under observation $P_{UO}$:

$$Retrieved\ ranked_{patents} = sort(R_{sd}, d_p \rightarrow final_{score}(d_p)) \quad (37)$$

where the $d_p \rightarrow final_{score}(d_p)$ explicitly denotes that each document $d_p$ is ranked based on its computed final score $final_{score}(d_p)$. This ranking ensures full recall, meaning all potentially relevant prior art is surfaced while favoring documents with strong semantic and conceptual alignment to the query patent.

Figures 2(a) through 2(d), along with Figures 3 and 4, illustrate the step-by-step cascading flow of information across the three phases of the FullRecall framework:

- Figures 2(a)-(d) illustrates the process of generating ranked noun phrases during Phase 1, highlighting key stages such as sentence filtering, noun phrase extraction, and ranking.
- Figure 3 illustrates how the top-ranked noun phrases are assembled into a structured search query in Phase 2 through intermediary intervention.
- Figure 4 illustrates the ranking of retrieved patent documents based on semantic similarity measures in Phase 3, resulting in the final set of relevant prior art documents.

This structured visual breakdown reinforces the logical progression of the proposed FullRecall framework, demonstrating how each component builds upon the previous component to achieve robust, scalable, and high-recall prior art retrieval; applicable to both patent examiners and inventors. Table 1 summarizes the mathematical notations introduced in the methodology, along with their corresponding definitions.

Table 1: List of notations and their descriptions

| Notation | Description |
| --- | --- |
| $P_{UO}$ | Patent under observation |
| $IPC_{P_{uo}} = \{IPC_1, IPC_2, \ldots, IPC_n\}$ | Set of $n$ IPC classifications for $P_{UO}$ |
| $S_{IPC_i} = \{S_{IPC_{i1}}, S_{IPC_{i2}}, \ldots S_{IPC_{im_i}}\}$ | $m_i$ subgroups for each IPC classification group |
| $D_{IPC_i}$ | Description associated with $IPC_i$ |
| $D_{S\_IPC_{ij}}$ | Description associated with subgroup $S_{IPC_{ij}}$ |
| $D$ | Set of all descriptions across all IPC classification groups and subgroups |
| $K_{phrases}$ | Key phrases from description $D_i$, where $D_i \in D$ |
| $K_{u\_phrases}$ | Unique key phrases from $D$ |
| $S_{P_{UO}} = \{s_1, s_2, \ldots, s_k\}$ | Set of sentences in $P_{UO}$ |
| $cos(kp_i, s_j)$ | Cosine similarity between the ith key phrase and the jth sentence of $P_{UO}$ |
| $S_{selected}$ | Set of selected sentences from $P_{UO}$ |

| | | |
|---|---|---|
| | $N_{phrases}$ | Set of noun phrases extracted from $S_{selected}$ |
| | $N_{u_{phrases}}$ | Set of unique noun phrases from $N_{phrases}$ |
| | $Em_{u_{phrases}}$ | Embeddings of noun phrases in $N_{u_{phrases}}$ |
| | $C$ | Set on $m$ clusters |
| | $G = (V, E)$ | A Graph, where each node $v_i \in V$ and an edge $e_i \in E$ |
| | $em_{c_j}$ | Centroid of cluster $c_j$ |
| | $em_{c_{graph}}$ | Global graph centroid |
| | $np_i$ | ith noun phrase where $np_i \in N_{u_{phrases}}$ |
| | $C_{np_i}$ | Connectivity score of $np_i$ |
| | $U_{np_i}$ | Uniqueness score of $np_i$ |
| | $Cc_{np_i}$ | distance of the noun phrase $np_i$ from its cluster centroid |
| | $Cg_{np_i}$ | distance of the noun phrase $np_i$ from the overall graph centroid |
| | $Uc_{np_i}$ | Intra-cluster uniqueness of the noun phrase $np_i$ |
| | $Ug_{np_i}$ | Cross-cluster distinctiveness of the noun phrase $np_i$ |
| | $P(np_i)$ | PageRank score of the noun phrase $np_i$ |
| | $B(np_i)$ | Betweenness centrality of the noun phrase $np_i$ |
| | $D(np_i)$ | Degree centrality of the noun phrase $np_i$ |
| | $S(np_i)$ | Semantic score of the noun phrase $np_i$ |
| | $R(np_i)$ | Final ranking score of the noun phrase $np_i$ |
| | $NP_{query}$ | Structured query using the top K noun phrases |
| | $S_d$ | Dataset consisting of patent documents from the same IPC classes as the query patent $P_{UO}$ |
| | $R_{sd}$ | Retrieved set from dataset $S_d$ |
| | $d_p \in R_{sd}$ | Retrieved patent document $d_p \in R_{sd}$ |
| | $e_{d_p}$ | Embedding of the retrieved patent document $d_p \in R_{sd}$ |
| | $e_{np_i}$ | Embedding of the noun phrase $np_i \in NP_{query}$ |
| | $\cos(e_{d_p}, e_{np_i})$ | Cosine similarity between the $e_{d_p}$ and $e_{np_i}$ |
| | $NP_{matches}(d_p)$ | Total count of noun phrases that matched each document $d_p \in R_{sd}$ |
| | $weighted_{score}(d_p)$ | Weighted score for a document $d_p \in R_{sd}$ |
| | $final_{score}(d_p)$ | Final score for each document $d_p \in R_{sd}$ |
| | $Retrieved\ ranked_{patents}$ | Retrieved ranked documents |

---

**Algorithm Phase 1:** Feature extraction - Ranked noun phrases extraction

**Input:** Patent under observation $P_{UO}$, full set of descriptions $D = \bigcup_{i=1}^{n}(\{D_{IPC_i}\} \cup \bigcup_{j=1}^{m_i}\{D_{S\_IPC_{ij}}\})$ extracted from IPC codes $IPC = \{IPC_1, IPC_2, \dots, IPC_n\}$ and their subgroup codes $\{S_{IPC_{i1}}, S_{IPC_{i2}}, \dots S_{IPC_{im_i}}\}$

**Output:** Ranked noun phrases, $R\left(N_{u_{phrases}}\right)$

1: $\textbf{\textit{execution of algorithm phase 1a}}$ {$\textbf{\textit{return}}$ $S_{selected}$ from $P_{UO}$} ← Full set of descriptions $D$
2: $\textbf{\textit{execution of algorithm phase 1b}}$ {$\textbf{\textit{return}}$ $N_{u_{phrases}}$ from $S_{selected}$} ← $S_{selected}$ from $P_{UO}$
3: $\textbf{\textit{execution of algorithm phase 1c}}$ {$\textbf{\textit{return}}$ Ranked noun phrases $R\left(N_{u_{phrases}}\right)$} ← $N_{u_{phrases}}$ from $S_{selected}$
4: $\textbf{\textit{return}}$ Ranked noun phrases $R\left(N_{u_{phrases}}\right)$

---

**Figure 2a:** Overall phase 1 of the algorithm - Feature extraction resulting in a ranked list of noun phrases

---

**Algorithm Phase 1a:** Selection of Key Sentence from $P_{UO}$

**Input:** Full set of descriptions $D$
**Output:** $S_{selected}$ from $P_{UO}$

1: $K_{u\_phrases}$ ← $extract\ n - gram\ (n \in 2,3)\ keyphrase$
2: $S_{P_{UO}} = \{s_1, s_2, \dots, s_k\}$ ← $extract\ sentences\ from\ P_{UO}$
3: $\textbf{\textit{for}}\ kp_i \in K_{u\_phrases}$ and $s_j \in S_{P_{UO}}$, $\textbf{do}$
4: $cos(kp_i, s_j) = \frac{kp_i \cdot s_j}{\|kp_i\|\|s_j\|}$ ← cosine similarity $kp_i \in K_{u\_phrases}$ and $s_j \in S_{P_{UO}}$
5: $\textbf{\textit{if}}\ sim(kp_i, s_j) > threshold$ ← Select $s_j$
6: $S_{selected}$ ← Add $s_j$ to selected sentences
7: $\textbf{\textit{return}}\ S_{selected}$

---

**Figure 2b:** Sub-phase (1a) of algorithm phase 1 - Selection of Key Sentences from $P_{UO}$

---

**Algorithm Phase 1b:** Extraction of unique noun phrases from $S_{selected}$ from $P_{UO}$

**Input:** $S_{selected}$ from $P_{UO}$
**Output:** Unique noun phrases $N_{u_{phrases}}$ from $S_{selected}$

1: $N_{phrases} = \{np_1, np_2, \dots, np_t\}$ ← $extract\ n - gram\ (n \in 2,3)\ noun\ phrases\ (S_{selected})$
2: $N_{u_{phrases}} = \{x | x \in N_{phrases}\}$ ← $remove\ duplicates(N_{phrases})$
3: $\textbf{\textit{return}}\ N_{u_{phrases}}$

---

**Figure 2c:** Sub-phase (1b) of algorithm phase 1 - Extraction of unique noun phrases

---

**Algorithm Phase 1c:** Ranking of noun phrases

**Input:** Unique noun phrases $N_{u_{phrases}}$

**Output:** Ranked noun phrases, $R\left(N_{u_{phrases}}\right)$

1: $C = c_1, c_2, \ldots, c_m \leftarrow \text{HSBSCAN}(N_{u_{phrases}})$
2: **Graph** $G = (V, E) \leftarrow C = c_1, c_2, \ldots, c_m$ and $N_{u_{phrases}}$
3: **for** $c_i \in C$ and $np_i \in N_{u_{phrases}}$, **do**
4: $\cos(em_i, em_j) = \frac{em_i \cdot em_j}{\|em_i\|\|em_j\|} \leftarrow c_i \in C$ and $np_i \in N_{u_{phrases}}$
5: **if** $\cos(em_i, em_j) > \tau$ **then**
6: $e_i \in E$ between nodes $v_i$ and $v_j$
7: **Semantic scoring** $S(np_i) = C_s$ and $U_s \leftarrow G = (V, E)$
8: **composite score** $\sum_{i=1}^{N} R(np_i) \leftarrow R(np_i) = \alpha . P(np_i) + \beta . D(np_i) + \delta . B(np_i) + (1 - \alpha - \beta - \delta) S(np_i)$
9: **Ranked noun phrases** $R\left(N_{u_{phrases}}\right) \leftarrow$ composite score $\sum_{i=1}^{N} R(np_i)$
10: **return** Ranked noun phrases $R\left(N_{u_{phrases}}\right)$

---

**Figure 2d:** Sub-phase (1c) of algorithm phase 1 - Ranking of noun phrases

---

**Intermediary intervention:** Conveyor information flow - Pipelining ranked noun-phrases to phase 2

**Input:** Ranked noun phrases $R\left(N_{u_{phrases}}\right)$

**Output:** query $Q_k$

$Q_k = f(NP_{query}) \leftarrow NP_{query} = \left\{np_i | np_i \in Top - K\left(R\left(N_{u_{phrases}}\right)\right), \; 12 \leq K \geq 16\right\}$

---

**Figure 3:** Intermediary intervention in the form of the query formulation

---

**Algorithm phase 2:** Full recall - Ranked retrieval of patent documents

**Input:** dataset $R_{sd}$, query $Q_k$

**Output:** Ranked retrieved documents $Retrieved\; ranked_{patents}$

1: $R_{sd} \leftarrow Search(S_d, Q_k)$
2: **for** $d_p \in R_{sd}$ and $np_i \in R\left(N_{u_{phrases}}\right)$
3: **if** $valid\left(e_{d_p}, e_{np_i}\right) = \begin{cases} 1, & \text{if } \cos\left(e_{d_p}, e_{np_i}\right) > \tau \\ 0, & \text{otherwise} \end{cases}$ ← cosine similarity
4: # of valid matches $NP_{matches}(d_p) = \sum_k valid(e_{d_p}, e_{np_k})$
5: **sort** $(NP_{matches}(d_p), descending)$
6: $weighted_{score}(d_p) = \sum_{k=1}^{NP_{matches}(d_p)} \frac{\cos\left(e_{d_p}, e_{np_i}^{(k)}\right)}{k}$
7: $final_{score}(d_p) = weighted_{score}(d_p) + \lambda \cdot NP_{matches}(d_p)$
8: $Retrieved\; ranked_{patents} = sort(R_{sd}, d_p \rightarrow final_{score}(d_p))$

---

**Figure 4:** Second phase of the algorithm - Retrieval of ranked relevant documents with Full Recall

## 3. Results and Discussions

Empirical experiments were conducted to evaluate the effectiveness of the proposed FullRecall framework in simulating real-world patent retrieval scenarios for both patent examiners and inventors. Patent examiners perform prior art searches for patent applications with IPC codes already pre-assigned, thus allowing targeted retrieval. On the other hand, inventors initiate prior art searches to determine patentability before filing a patent application. Thus, they must first infer the appropriate classifications. This can be done by reviewing similar patents or employing AI-driven tools such as WIPO IPCCAT or BERT for Patents. Once IPC codes are assigned, both user types follow the same retrieval methodology.

The objective is to identify and rank all the relevant prior art from a huge collection of prior art documents. Five patents under observation, denoted as $P_{1\_UO}$, $P_{2\_UO}$, $P_{3\_UO}$, $P_{4\_UO}$ and $P_{5\_UO}$, were randomly selected across diverse IPC domains and treated as if they were new patent filings requiring exhaustive prior art searches. To evaluate the effectiveness of the proposed FullRecall retrieval process, patent examiner citations associated with each patent serve as ground truth, providing a reliable benchmark for assessing retrieval performance. These citations, obtained from MineSoft PatBase® [46], represent the official prior art identified during prosecution.

Table 2 lists the relevant examiner-cited documents for each patent under observation: $P_{1\_UO}$, $P_{2\_UO}$, $P_{3\_UO}$, $P_{4\_UO}$ and $P_{5\_UO}$. $P_{1\_UO}$ and $P_{5\_UO}$ are associated with 10 and 7 citations, respectively, indicating strong prior art relationships, whilst $P_{2\_UO}$, $P_{3\_UO}$ and $P_{4\_UO}$ are associated with 4, 6 and 5 citations, respectively, indicating moderate levels of prior art connections.

**Table 2:** Patents Under Observation and Their Relevant Examiner-cited Prior Art

| Patent_under_observation | | examiner_citations | Number_of_examiner_citations |
|---|---|---|---|
| $P_{1\_UO}$ | US2019053227A1 | EP3454474 A1, US20180262242 A1, US20180324768 A1, US20190014481 A1, US20190053227 A1, US20190363843 A1, US20200037297 A1, US20200059867 A1, US20220279537 A1, US20230098368 A1 | 10 |
| $P_{2\_UO}$ | US20200124681A1 | WO2017038387 A1, US20140302355 A1, EP3346542 A1, US20120176096 A1' | 4 |
| $P_{3\_UO}$ | US10825607 B2 | US20130192860 A1, US20090192667 A1, US6072265 A, US4945269 A, US4631959 A, US4075603 A | 6 |
| $P_{4\_UO}$ | US12168226 B2 | US20140066318 A1, US20130116128 A1, US20130053256 A1, US20120245053 A1, CA2473308 A1 | 5 |
| $P_{5\_UO}$ | US8976843 B2 | US20050063345 A1, US20040246998 A1, US8588326 B2, US20040131007 A1, US8102832 B2, US20070097927 A1, US7379742 B2 | 7 |

**Table 3:** Patents under observation and their corresponding IPC classification codes

| Patent_under_observation | | IPC Classification codes |
|---|---|---|
| $P_{1\_UO}$ | US2019053227A1 | H04L25/03, H04L5/00, H04W72/04, H04W74/08 |
| $P_{2\_UO}$ | US20200124681A1 | B60L58/12, B60L58/16, G01R31/374, G01R31/387, G01R31/389, G01R31/392, H01M10/42, H01M10/48 |
| $P_{3\_UO}$ | US10825607 B2 | B60C23/00, H01F27/02, H01F38/14, H01F38/18, H02J50/12, F01D17/10, F02C6/08, F04D27/02, H02J50/80, H04B5/00 |
| $P_{4\_UO}$ | US12168226 B2 | C12Q1/6816, B01L3/00, C12Q1/68, C12Q1/6851, C12Q1/6853, C40B20/02, C40B20/04, C40B40/06, C40B40/10, B01L7/00 |
| $P_{5\_UO}$ | US8976843 B2 | H04B1/707, H04B7/06, H04J13/00, H04L5/00, H04L25/02, H04L27/26, H04J13/16, H04L25/03 |

Each patent under observation $P_{i\_UO}$ is associated with one or more IPC codes $IPC_{P_{i\_uo}}$ [46]. These codes not only help define the technological domain of each invention but also, within the context of the FullRecall framework, serve as structured filters to narrow down the search space for prior art retrieval. Table 3 illustrates the IPC codes associated to each patent under observation, $P_{1\_UO}$, $P_{2\_UO}$, $P_{3\_UO}$, $P_{4\_UO}$ and $P_{5\_UO}$; with each IPC code tagged with standardized descriptions that characterize the technology domain, highlight problem contexts, and articulate the technical innovations being claimed [47].

To build a structured prior knowledge base $D$, the descriptions corresponding to each assigned IPC class, including subclasses, groups, and subgroups, are aggregated for every patent under observation. This knowledge base provides domain-specific context that guides downstream analysis, including the extraction of key technical phrases and the evaluation of semantic relevance during retrieval.

The retrieval process begins with the first phase of the framework, which consists of three sub-phases. In the first sub-phase, the knowledge database $D$ associated with a given patent under observation $P_{i\_UO}$ is processed using YAKE to generate a set of unique key phrases $K_{u\_phrases}$ [48]. YAKE is a widely used unsupervised keyword extraction method that identifies statistically significant phrases within a document based on features such as term frequency and co-occurrence, without requiring any reference corpus [48]; making it suitable for single-document processing. The resulting phrases are ranked and serve as input for formulating precise search queries.

Table 4 lists the unique key phrases generated for the five (5) patents under observation: $P_{1\_UO}$, $P_{2\_UO}$, $P_{3\_UO}$, $P_{4\_UO}$ and $P_{5\_UO}$. The diversity of extracted terms reflects the technological breadth across the selected patents, from wireless communication and electric propulsion systems to biochemical libraries and laboratory processes. For instance, $P_{1\_UO}$ and $P_{5\_UO}$, both falling under electrical engineering domains, yielded frequent references to communication techniques, transmission systems, and modulation schemes. Meanwhile, $P_{4\_UO}$, a biochemical technology patent, generated phrases related to genetic engineering, nucleotides, and enzymatic processes. These results indicate that the YAKE-based extraction process effectively captures domain-specific technical terminology that can later support targeted query formulation and semantic relevance scoring.

**Table 4:** Patents under observation and their extracted key phrases from IPC descriptions

| Patent_under_observation | | | | | | | | | |
|---|---|---|---|---|---|---|---|---|---|
| $P_{1\_UO}$ (US2019053227A1) | | $P_{2\_UO}$ (US20200124681A1) | | $P_{3\_UO}$ (US10825607 B2) | | $P_{4\_UO}$ (US12168226 B2) | | $P_{5\_UO}$ (US8976843 B2) | |
| electricity | signalling resource management | physics | electric elements | electricity | product engine plants | chemistry | label | electricity | |
| electric communication technique | direction wireless link | measuring | process means | electric elements | gas turbine plants | metallurgy | associated library members | electric communication technique | |
| wireless communication networks | random access procedures | testing | energy electrical energy | magnets | jet propulsion plants | biochemistry | decoding processes | transmission | |
| wireless channel access | | measuring electric variables | secondary cells | inductances | jet propulsion plants | beer | libraries per | details transmission systems | |
| non scheduled access | | measuring magnetic variables | maintenance secondary cells | transformers | gas turbine plants | spirit | arrays | transmission systems | |
| carrier sensing | | testing electric properties | secondary half cells | selection materials | gas turbine plants | wine | mixtures | single one groups | |
| sense multiple access | | locating electric faults | process | magnetic properties | plants special use | vinegar | containing organic compounds | medium used transmission | |
| transmission | | arrangements electrical testing | means | adaptations transformers | working fluid apparatus | microbiology | libraries containing nucleotides | spread spectrum techniques | |
| radio transmission systems | | electrical condition accumulators | combined arrangements measuring | specific applications functions | mechanical power output | enzymology | libraries containing polynucleotides | direct sequence modulation | |
| radiation field | | electric batteries | testing condition cells | rotary transformers | compressed gas | mutation | libraries containing derivatives | radio transmission systems | |
| diversity systems | | capacity state charge | indicating condition cells | performing operations | gas turbine compressor | genetic engineering | libraries containing peptides | radiation field | |
| multi antenna systems | | performing operations | level density electrolyte | transporting | displacement machines liquids | processes involving enzymes | libraries containing polypeptides | diversity systems | |
| independent antennas | | transporting | | vehicles | liquids elastic fluids | nucleic acids | operations | multi antenna systems | |
| comprising multiple antennas | | vehicles general | | vehicles tyres | positive displacement pumps | microorganisms | physical apparatus laboratory | using multiple antennas | |
| uplink diversity | | electrically propelled vehicles | | tyre inflation | control | compositions test papers | heating apparatus | spaced independent antennas | |
| transmitting station | | electrically propelled vehicles | | tyre changing | regulation pumps | processes preparing compositions | cooling apparatus | independent antennas | |
| transmission digital information | | brake systems vehicles | | inflatable elastic bodies | pumping installations | microbiological enzymological processes | heat insulating devices | transmitting station | |
| telegraphic communication | | electrically propelled vehicles | | arrangements related tyres | adapted elastic fluids | involving nucleic acids | | multiplex communication | |
| errors information received | | electrically propelled vehicles | | devices measuring | surge control | hybridisation assays | | transmission digital information | |
| use transmission path | | circuit arrangements monitoring | | signalling | materials magnetic properties | characterised detection means | | telegraphic communication | |
| using type signal | | controlling batteries | | controlling | details transformers inductances | performing operations | | use transmission path | |

| | | | | | |
|---|---|---|---|---|---|
| duplex | fuel cells | tyre pressure temperature | casings | transporting | transmission path |
| power management | adapted electric vehicles | adapted mounting vehicles | specific applications functions | physical processes | baseband systems |
| transmission power control | monitoring batteries | inflating devices vehicles | inductive couplings | chemical processes | details |
| according specific parameters | state charge | pumps | generation electric power | apparatus general | modulated carrier systems |
| rate quality service | monitoring controlling batteries | tanks | conversion electric power | chemical laboratory apparatus | multi frequency codes |
| user profile | battery ageing | tyre cooling arrangements | distribution electric power | containers laboratory use | code allocation |
| mobile speed | number charging cycles | mechanical engineering | circuit arrangements | dishes laboratory | shaping networks transmitter |
| priority network state | state health | lighting | distributing electric power | laboratory glassware | shaping networks networks receiver |
| performed specific situations | electrical condition accumulators | heating | storing electric energy | droppers | |
| performed systems time | correcting measurement temperature | weapons | systems wireless supply | acid amplification reactions | |
| space | correcting measurement ageing | blasting | inductive coupling | quantitative amplification | |
| frequency | testing electric properties | machine | resonant type | modified primers | |
| polarisation diversity | locating electric faults | engines | involving exchange data | modified templates | |
| local resource management | arrangements testing | engine plants | supply electric power | combinational technology | |
| wireless resource allocation | arrangements measuring battery | steam engines | transmitting devices | combinational chemistry | |
| type allocated resource | accumulator variables | positive displacement machines | receiving devices | libraries | |
| resources time domain | hour charge capacity | positive displacement engines | electric communication technique | chemical libraries | |
| wireless traffic scheduling | tested provided elsewhere | steam turbines | transmission | identifying library members | |
| traffic onto schedule | measuring internal impedance | controlling varying flow | field transmission systems | physical location support | |
| scheduled allocation | internal conductance | final actuators | inductive transmission systems | physical location support | |
| multiplexing flows | related variables | combustion engines | capacitive transmission systems | physical location substrate | |
| downlink data flows | electricity | hot gas | | members means tag | |

In the second sub-phase of the first phase, the unique key phrases $K_{u\_phrases}$, extracted from IPC descriptions, are used to query the corresponding patent document under observation $P_{1\_UO}$, $P_{2\_UO}$, $P_{3\_UO}$, $P_{4\_UO}$ and $P_{5\_UO}$. Sentences that exhibit a cosine similarity greater than a predetermined threshold $\tau$ with any of the key phrases are selected, forming a set of semantically relevant sentences $S_{selected}$. From this sentence set, distinct noun phrases $N_{u\_phrases}$ are extracted using spaCy [49], a widely used NLP library. spaCy's noun phrase chunking functionality is applied to identify linguistically structured noun-based expressions, which are essential since over 90% of technical terms in patent literature are noun phrases [44]. These noun phrases serve as condensed, high-information descriptors of the technical content in each patent and are retained for further ranking and query formulation.

In the third and final sub-phase of the first phase, the noun phrases $N_{u\_phrases}$ are then clustered using HDBSCAN [50], an unsupervised machine learning technique that extends DBSCAN [51] by enabling variable-density clusters and identifying outliers through a hierarchical process [50]. Clustering effectively organizes the noun phrases into semantically coherent group. A graph $G = (V, E)$ is then constructed for each patent under observation $P_{1\_UO}$, $P_{2\_UO}$, $P_{3\_UO}$, $P_{4\_UO}$ and $P_{5\_UO}$, with each node $v_i \in V$ representing a noun phrase.

Edges $e_{ij} \in E$ are instantiated between nodes if the cosine similarity between their embeddings exceed a specified threshold, indicating a level of strength to their semantic relationship. This graph-based representation provides a structured visualization of the relationships between key technical terms, supporting more effective query formulation and prior art retrieval. Visual representations of noun phrase graphs for the five patents under observation $P_{1\_UO}$, $P_{2\_UO}$, $P_{3\_UO}$, $P_{4\_UO}$ and $P_{5\_UO}$, are illustrated in Figures 3 through 7, respectively; offering an intuitive overview of the extracted terminology networks for each patent.

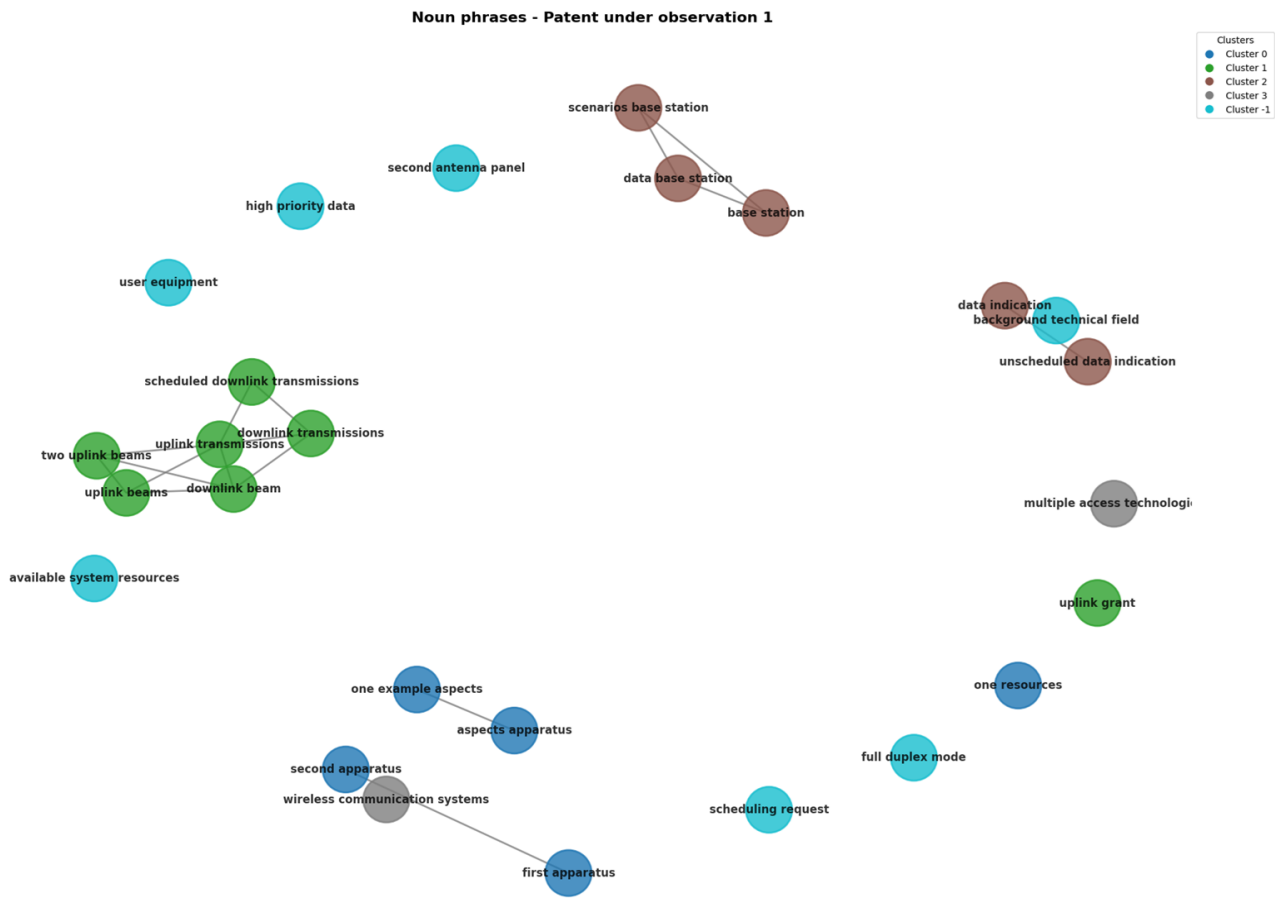

**Figure 3:** Graphical representation of noun phrases for $P_{1\_UO}$

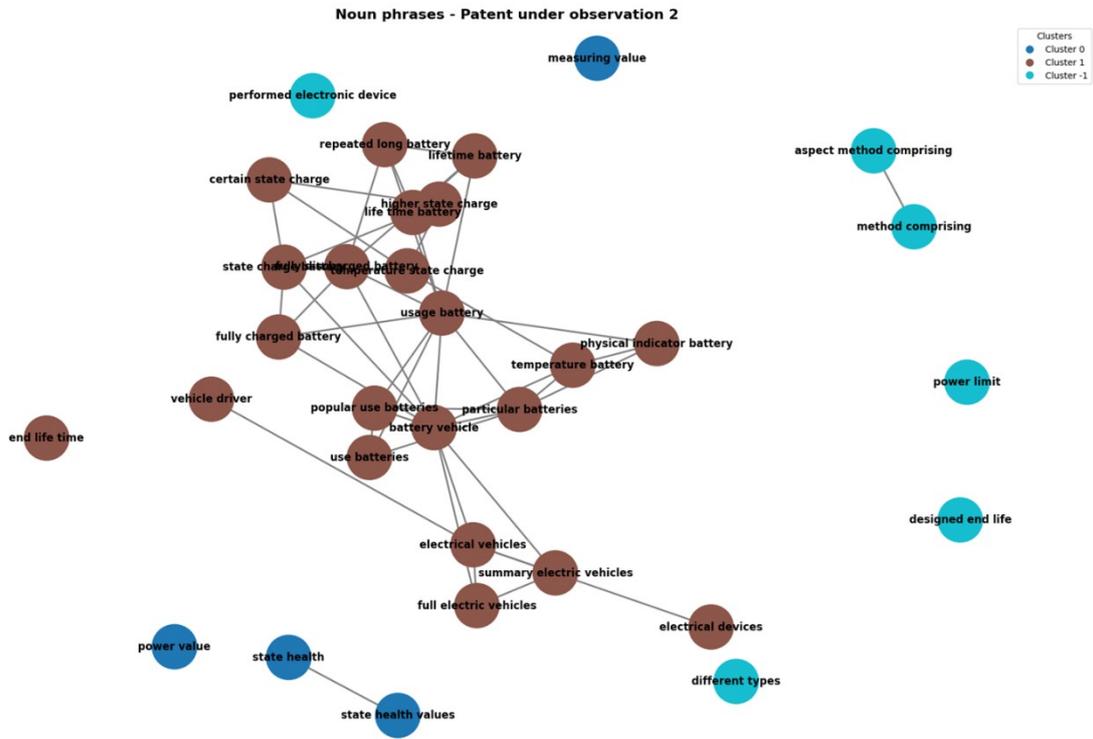

**Figure 4:** Graphical representation of noun phrases for $P_{2\_UO}$

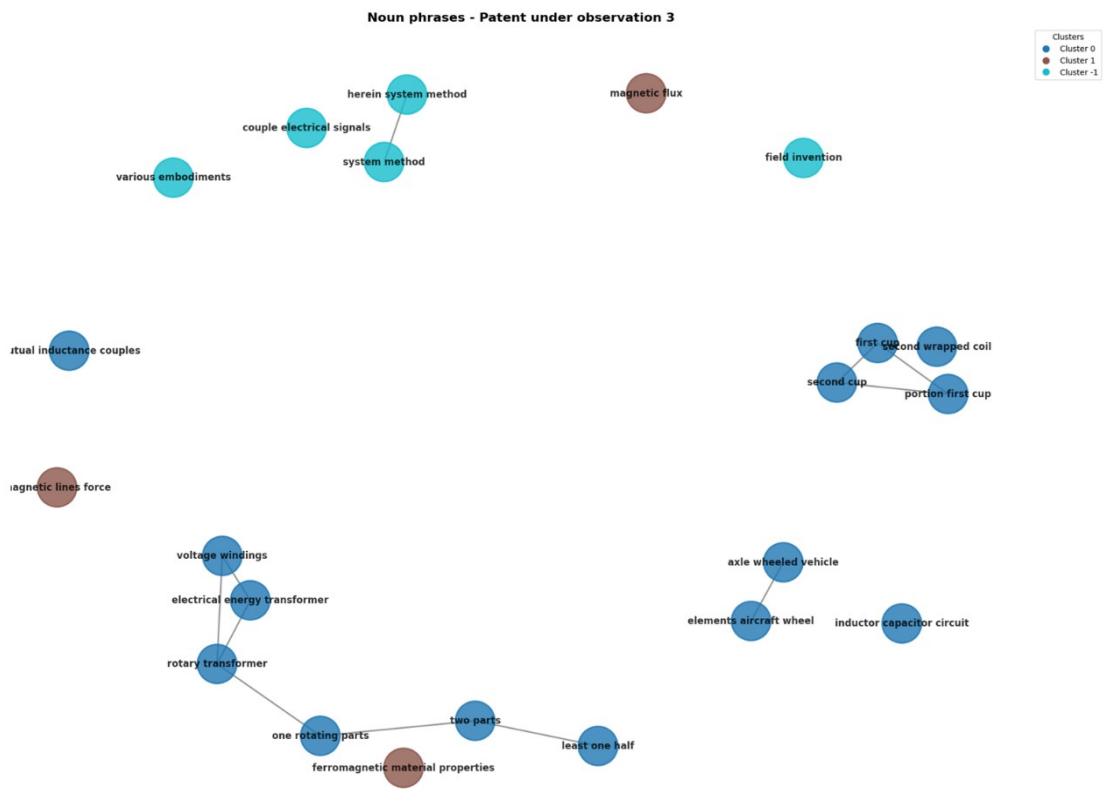

**Figure 5:** Graphical representation of noun phrases for $P_{3\_UO}$

**Figure 6:** Graphical representation of noun phrases for $P_{4\_UO}$

**Figure 7**: Graphical representation of noun phrases for $P_{5\_UO}$

Within each graph representation $G$, every node representing a noun phrase $np_i \in N_{u_{phrases}}$, is assigned a semantic score $S(np_i)$, derived from two key components: connectivity score $C_{np_i}$ and uniqueness score $U_{np_i}$, as defined in equations (18) and (21), respectively. These scores are computed both at the intra-cluster (local) and graph-wide (global) levels to provide a comprehensive assessment of the significance of each noun phrase. The connectivity score $C_{np_i}$ includes an intra-cluster score $Cc_{np_i}$, which quantifies the distance from a node to the centroid of its cluster, and a graph-level score $Cg_{np_i}$, representing the distance from the node to the centroid of the entire graph. Small $Cc_{np_i}$ and $Cg_{np_i}$ values indicate closer alignment with the cluster's central concept and the overall theme of the document, respectively. Simultaneously, the uniqueness score $U_{np_i}$ consists of $Uc_{np_i}$ measuring the cosine similarity between the node and other nodes within its respective cluster, and $Ug_{np_i}$ measuring its similarity to centroids of other clusters. Small $Uc_{np_i}$ and $Ug_{np_i}$ values imply stronger semantic independence; identifying the noun phrase as more distinct and informative, whilst large values identify the noun phrase as generic and hence, conceptually redundant within or across clusters. Semantic score $S(np_i)$ is then computed as the average of $C_{np_i}$ and $U_{np_i}$ as given in equation (24), integrating both centrality and distinctiveness of each phrase.

To further refine phrase relevance, graph-based centrality measures are computed, including PageRank $P(np_i)$, Degree Centrality $D(np_i)$ and Betweenness Centrality $B(np_i)$ [52]. PageRank $P(np_i)$, originally developed for ranking web pages, is used here to assess the relative importance of a noun phrase based on the overall link structure of the graph. A high PageRank indicates that a node is connected to other highly influential nodes, suggesting its centrality in the flow of technical concepts. Degree Centrality $D(np_i)$, defined in equation (26), captures how well-connected a node is within the graph; with higher values indicating broader relevance and frequent referencing across the document. On the other hand, Betweenness Centrality $B(np_i)$, defined in equation (27), reflects the role of the node mediating information flow within the graph, with higher values highlighting phrases that bridge key technical concepts.

The final relevance score $R(np_i)$ for each noun phrase is obtained by aggregating PageRank $P(np_i)$, semantic score $S(np_i)$, degree centrality $D(np_i)$, and betweenness centrality $B(np_i)$, as given in equation (28). This ranking function ensures that only the most central, distinctive, and semantically informative phrases are prioritized. The resulting list $R(np_i)$ represents the output of the third sub-phase in the first phase of the algorithm and forms the input for the next stage of query generation and retrieval.

In the second phase, the ranked noun phrases $R(np_i)$ are compiled into a structured ranking list for each patent under observation: $P_{1\_UO}$, $P_{2\_UO}$, $P_{3\_UO}$, $P_{4\_UO}$ and $P_{5\_UO}$. To explore the influence of query length on retrieval performance and prior art coverage, a subset of the ranked noun phrases is selected by varying the number of top-$k$ phrases, where $12 \leq K \geq 16$. For each patent $P_{1\_UO}$, $P_{2\_UO}$, $P_{3\_UO}$, $P_{4\_UO}$ and $P_{5\_UO}$, and chosen k value, a corresponding query $Q_k$ is formulated. This well-structured query $Q_k$ is the core output generated of phase 2 and is specifically structured to systematically probe various sections of a patent document. In the current implementation, the query $Q_k$ targets five specific sections of a patent document: title, abstract, claims, first claim, and independent claims. Each query $Q_k$ is composed of highly ranked noun phrases from Phase 1, ensuring semantic precision and maximizing technical relevance in identifying prior art. By tailoring $Q_k$ to these important sections, the FullRecall framework enhances the recall and precision of the prior art search process. Table 5 highlights the top-ranked noun phrases used in constructing query $Q_k$ for each of the patents under observation: $P_{1\_UO}$, $P_{2\_UO}$, $P_{3\_UO}$, $P_{4\_UO}$ and $P_{5\_UO}$; providing insight into the core technical themes prioritized during query formulation.

**Table 5:** Patents Under Observation and Their Ranked Noun Phrases in Descending Order

| Patent_under_observation | | | | |
|---|---|---|---|---|
| $P_{1\_UO}$ (US2019053227A1) | $P_{2\_UO}$ (US20200124681A1) | $P_{3\_UO}$ (US10825607 B2) | $P_{4\_UO}$ (US12168226 B2) | $P_{5\_UO}$ (US8976843 B2) |
| uplink transmissions | battery vehicle | rotary transformer | amplification products | plurality antennas transmitter |
| downlink beam | usage battery | one rotating parts | template nucleic acids | wherein transmitter |
| downlink transmissions | electrical vehicles | two parts | amplifying nucleic acids | pilot group |
| uplink beams | state charge battery | first cup | oligonucleotide array | pilots sub carrier |
| two uplink beams | fully discharged battery | electrical energy transformer | oligonucleotide tile array | least two transmitters |
| base station | particular batteries | second cup | correlating amplification product | embodiments receiver |
| data base station | temperature state charge | portion first cup | rise amplification products | least one antenna |
| scenarios base station | temperature battery | voltage windings | devices methods | plurality clusters |
| one example aspects | popular use batteries | herein system method | spatially specific manner | plurality instances cluster |
| data indication | lifetime battery | system method | second barcode | two antennas method |
| wireless communication systems | fully charged battery | axle wheeled vehicle | laborious steps | plurality instances |

| unscheduled data indication | repeated long battery | elements aircraft wheel | plurality amplification primers | embodiments plurality clusters |
| first apparatus | summary electric vehicles | magnetic flux | porous material | plurality contiguous groups |
| second apparatus | full electric vehicles | second wrapped coil | cdna molecules | ofdm symbols |
| scheduled downlink transmissions | certain state charge | ferromagnetic material properties | cellular sample | multiple ofdm symbols |
| scheduling request | higher state charge | field invention | active process | plurality ofdm symbols |

In the final phase of the FullRecall framework, a candidate dataset $S_d$ is created each patent under observation $P_{1\_UO}$, $P_{2\_UO}$, $P_{3\_UO}$, $P_{4\_UO}$ and $P_{5\_UO}$, containing prior art documents from the same IPC classes as the respective patent under observation. These datasets also include the examiner-cited patents identified earlier in Table 2, ensuring that relevant prior art is captured.

Each candidate dataset $S_d$ is preprocessed using Python libraries including NLTK, unicodedata, and re libraries, to prepare and standardize the text for further analysis. Texts from the Abstract, Claims, First Claim, and Independent Claims sections are transformed into a normalized format: lowercased, stripped of punctuation and non-ASCII characters, and cleaned of diacritical marks. Regular expressions (re) are employed to remove non-alphabetic symbols and redundant whitespace. Additionally, common stopwords are filtered out using NLTK's standard stopword list, reducing linguistic noise and enhancing semantic clarity.

After preprocessing, the sizes of each candidate dataset $S_d$ are:

- $S_d$ for $P_{1_{UO}}$ contains 4,795 documents;
- $S_d$ for $P_{2_{UO}}$ contains 5,004 documents;
- $S_d$ for $P_{3_{UO}}$ contains 5,006 documents;
- $S_d$ for $P_{4_{UO}}$ contains 5,505 documents; and
- $S_d$ for $P_{5_{UO}}$ contains 5,507 documents.

For each patent under observation, $P_{1\_UO}$, $P_{2\_UO}$, $P_{3\_UO}$, $P_{4\_UO}$ and $P_{5\_UO}$ and associated query $Q_k$ with $12 \leq k \leq 16$, the query $Q_k$ is executed over the corresponding candidate dataset $S_d$ to generate a retrieved result set $R_{sd}$. This set consists of documents most relevant to the ranked noun phrases in the query. Multiple values of $k$ are tested to assess retrieval performance and coverage. Table 6 presents the retrieval outcomes for each patent under observation across a range of query lengths, defined by the number of ranked noun phrases $k$. For each patent-specific dataset $S_d$, the corresponding retrieved result set $R_{sd}$ and the number of examiner-cited patents captured within it are reported.

Across all patents under observation, increasing the value of $k$ generally improves retrieval performance. Notably, for $P_{1\_UO}$, the number of examiner citations retrieved jumps significantly from 3 (at $k = 12 \rightarrow 14$) to 9 and 10 at $k = 15$ and $k = 16$, respectively. A similar pattern is observed for $P_{5\_UO}$, where only 1 citation is found at $k = 12$, but all 7 examiner-cited documents are retrieved when $k \geq 15$. This demonstrates the importance of sufficient phrase coverage in constructing effective queries. While higher $k$ values generally yield larger result sets, the increase in relevant citation retrieval is not always linear. For instance, $P_{2\_UO}$ maintains stable performance between $k = 12$ and $k = 16$, retrieving all 4 citations even at lower query lengths. This suggests that for some patents, key phrases ranked early already encapsulate the core technical content. Empirical observations indicate that query lengths of approximately 15–16 phrases consistently succeed in retrieving all target documents, suggesting this range offers an optimal balance between precision and recall. All five patents under observation achieve full citation recall at this range, validating the ranked noun phrase approach employed during query formulation. Overall, Table 6 confirms that adaptive query length tuning, guided by semantic and graph-based ranking, substantially enhances prior art retrieval effectiveness across varied technological domains.

**Table 6:** Patents under observation and the corresponding retrieved result set across a range of k values (# of key phrases)

| Patent_under_observation | | Dataset ($S_d$) | Value of k (# of key phrases) | Retrieved Result set ($R_{sd}$) | Number of examiner_citations found in the resulted set ($R_{sd}$) | Number of examiner_citations |
|---|---|---|---|---|---|---|
| $P_{1_{UO}}$ | US2019053227A1 | 4795 | 12 | 824 | 3 | 10 |
| | | | 13 | 553 | 3 | |
| | | | 14 | 553 | 3 | |
| | | | 15 | 2366 | 9 | |
| | | | 16 | 2342 | 10 | |
| $P_{2_{UO}}$ | US20200124681A1 | 5004 | 12 | 3494 | 3 | 4 |
| | | | 13 | 3494 | 3 | |
| | | | 14 | 3497 | 3 | |
| | | | 15 | 3635 | 4 | |
| | | | 16 | 3635 | 4 | |
| $P_{3\_UO}$ | US10825607 B2 | 5006 | 12 | 965 | 4 | 6 |
| | | | 13 | 1086 | 5 | |
| | | | 14 | 1086 | 5 | |
| | | | 15 | 1217 | 6 | |
| | | | 16 | 1217 | 6 | |
| $P_{4\_UO}$ | US12168226 B2 | 5505 | 12 | 1205 | 2 | 5 |
| | | | 13 | 1205 | 2 | |
| | | | 14 | 1206 | 2 | |
| | | | 15 | 2527 | 4 | |
| | | | 16 | 2426 | 5 | |
| $P_{5\_UO}$ | US8976843 B2 | 5507 | 12 | 329 | 1 | 7 |
| | | | 13 | 675 | 3 | |
| | | | 14 | 1448 | 6 | |
| | | | 15 | 1879 | 7 | |
| | | | 16 | 1879 | 7 | |

After retrieval, each patent documents in the retrieved result set $R_{sd}$ corresponding to the patents under observation: $P_{1\_UO}$, $P_{2\_UO}, P_{3\_UO}, P_{4\_UO}$ and $P_{5\_UO}$, undergoes a scoring and ranking process to prioritize the most relevant prior art. The independent claim section of each document is analyzed, as independent claims typically encapsulate the core inventive concepts of a patent.

To assess relevance, each document is compared against the ranked noun phrases previously extracted from its respective patent under observation. The comparison process incorporates both semantic similarity and the relative importance of each noun phrase:

- Each document is evaluated against the ranked noun phrase list of the corresponding $P_{i\_UO}$.
- Cosine similarity is computed between the document's independent claim and each noun phrase.
- These similarity scores are weighted by the noun phrase's rank, ensuring that higher-ranked (i.e., more semantically and structurally significant) phrases contribute more to the final score.
- The final document score integrates the weighted similarity and the count of matching noun phrases, as formalized in equation (36).

Table 7 showcases the five patents under observation: $P_{1\_UO}$, $P_{2\_UO}, P_{3\_UO}, P_{4\_UO}$ and $P_{5\_UO}$, along with their respective datasets $S_d$, the size of the retrieved result sets $R_{sd}$, and the final ranked sets $Retrieved\ ranked_{patents}$ that include all target patents (examiner citations). A notable observation is the significant reduction in the size of the final ranked sets compared to the initial retrieved result sets. For instance, for $P_{2\_UO}$, the result set is narrowed from 3635 to 895 documents, while still successfully including all four examiner-cited patents. Similar improvements are observed across all five test cases. This shows that the FullRecall framework not only retrieves the correct prior art but also ranks them effectively, significantly reducing the number of documents a patent examiner would need to review; thus improving both precision and interpretability.

This scoring mechanism ensures that documents containing highly ranked and semantically aligned noun phrases are prioritized. The final ranked result set successfully includes all examiner-cited patents, validating the effectiveness of the FullRecall framework. As examiner citations are considered authoritative by domain experts during the patent examination process, their presence in the final ranked set serves as strong empirical validation of the method. The results highlight the practical utility of FullRecall in supporting real-world patent search and examination workflows.

**Table 7:** Overview of Retrieved and Final Ranked Sets for each Patent under Observation: $P_{1\_UO}$, $P_{2\_UO}, P_{3\_UO}, P_{4\_UO}$ and $P_{5\_UO}$, demonstrating FullRecall Performance

| Patent_under_observation | | Dataset ($S_d$) | Retrieved Result set ($R_{sd}$) | Ranked set with FullRecall | Target patents (examiner_citations) | Number of target patents found |
|---|---|---|---|---|---|---|
| $P_{1_{UO}}$ | US2019053227A1 | 4795 | 2342 | 1660 | EP3454474 A1, US20180262242 A1, US20180324768 A1, US20190014481 A1, US20190053227 A1, US20190363843 A1, US20200037297 A1, US20200059867 A1, US20220279537 A1, US20230098368 A1 | 10 |
| $P_{2_{UO}}$ | US20200124681A1 | 5004 | 3635 | 895 | WO2017038387 A1, US20140302355 A1, | 4 |

|  |  |  |  |  | EP3346542 A1, US20120176096 A1 |  |
|---|---|---|---|---|---|---|
| $P_{3_{UO}}$ | US10825607 B2 | 5006 | 1217 | 1030 | US20130192860 A1, US20090192667 A1, US6072265 A, US4945269 A, US4631959 A, US4075603 A | 6 |
| $P_{4\_UO}$ | US12168226 B2 | 5505 | 2426 | 1473 | US20140066318 A1, US20130116128 A1, US20130053256 A1, US20120245053 A1, CA2473308 A1 | 5 |
| $P_{5\_UO}$ | US8976843 B2 | 5507 | 1879 | 1568 | US20050063345 A1, US20040246998 A1, US8588326 B2, US20040131007 A1, US8102832 B2, US20070097927 A1, US7379742 B2 | 7 |

## 4. Performance Comparison

To further demonstrate the effectiveness of the proposed FullRecall framework, a comparative evaluation is conducted against the two established baselines: High Recall Retrieval with Relevance feedback (HRR2) [1] and Retrieval using Query Reformulation and Relevance Classification (ReQ-ReC) [2], using the same patents under observation: $P_{1\_UO}$, $P_{2\_UO}$, $P_{3\_UO}$, $P_{4\_UO}$ and $P_{5\_UO}$. HRR2 is a two-step high recall retrieval technique aimed at extracting relevant documents from large-scale datasets [1]. HRR2 takes the value of k as input, indicating the number of relevant documents to retrieve. Although the algorithm is designed to stop once all k relevant documents are found, in this implementation, HRR2 was executed over the entire dataset to avoid premature termination before identifying any or few relevant documents. In contrast, ReQ-ReC iteratively alternates between query expansion and classifier refinement to progressively improve retrieval quality [2]. Typically, ReQ-ReC [2] limits its results to a top-n ranked list (e.g. top 10, 20, or 30 documents, which may constrain recall. To ensure a meaningful comparison, the retrieval boundary for ReQ-ReC was expanded to match the point at which FullRecall achieved 100% recall for each query patent [1]. For instance, in the case of query patent $P_{1\_UO}$ in FullRecall setting, all relevant documents were retrieved within the top 1660 documents of the ranked list. Therefore, the value of n for ReQ-ReC was set to 1660 to evaluate how many relevant documents ReQ-ReC could identify within that range. Similarly for query patent $P_{2\_UO}$, $P_{3\_UO}$, $P_{4\_UO}$ and $P_{5\_UO}$ the value of n for ReQ-ReC was set to 895, 1030, 1473 and 1568 to determine the count of relevant documents ReQ-ReC could retrieve within that range. The performance of FullRecall is evaluated alongside HRR2 [1] and ReQ-ReC [2] to determine its relative effectiveness in retrieving relevant prior art.

Table 8 presents a side-by-side comparison of FullRecall with both HHR2 and ReQ-ReC across all five patents under observation. For each approach, the table highlights the number of retrieved documents, and the number of examiner-cited target patents found within the retrieved document set.

**Table 8:** Comparison of retrieved results with the proposed FullRecall and baseline approach HRR2 [1]

| Patent under observation | | Target patents (examiner_citations) | FullRecall (Retrieved documents) | FullRecall - # of target patents found | HRR2 (Retrieved documents) | HRR2 - # of target patents found | ReQ-ReC (Retrieved documents) | ReQ-ReC - # of target patents found |
|---|---|---|---|---|---|---|---|---|
| $P_{1_{UO}}$ | US2019053227A1 | 10 | 1660 | 10 | 140 | 1 | 1660 | 5 |
| $P_{2_{UO}}$ | US20200124681A1 | 4 | 895 | 4 | 175 | 1 | 895 | 1 |
| $P_{3_{UO}}$ | US10825607 B2 | 6 | 1030 | 6 | 169 | 2 | 1030 | 0 |
| $P_{4_{UO}}$ | US12168226 B2 | 5 | 1473 | 5 | 193 | 0 | 1473 | 0 |
| $P_{5_{UO}}$ | US8976843 B2 | 7 | 1568 | 7 | 184 | 1 | 1568 | 0 |

Across all five patents under observation, FullRecall consistently retrieved all examiner-cited target patents. For $P_{1_{UO}}$, FullRecall retrieved a total of 1,660 documents containing all the 10 target patents. In comparison, HRR2 retrieved only 140 documents but capturing just 1 of the 10 target patents, whilst ReQ-ReC retrieved the same number of documents as FullRecall but only successfully retrieved only 5 target patents. A similar pattern emerges for $P_{2_{UO}}$: FullRecall captured all 4 target patents within the 895 retrieved documents, while both HRR2 and ReQ-ReC captured only 1 out of the 4 target documents. For $P_{3_{UO}}$, $P_{4_{UO}}$, and $P_{5_{UO}}$, FullRecall successfully retrieved all target patents, achieving complete recall in each case. It retrieved 1,030 documents for $P_{3_{UO}}$, 1,473 documents for $P_{4_{UO}}$, and 1,568 documents for $P_{5_{UO}}$, encompassing all 6, 5, and 7 examiner-cited patents, respectively. In comparison, HRR2, while retrieving fewer documents (169 documents for $P_{3_{UO}}$, 18 documents for $P_{4_{UO}}$, and 23 documents for $P_{5_{UO}}$), managed to capture only 2, 0, and 1 target patents, respectively. ReQ-ReC, despite operating within the same expanded retrieval threshold, failed to retrieve any target patents in these three cases.

These findings highlight the superior recall performance of FullRecall, particularly in contexts where missing even a single relevant document is unacceptable. The lower number of documents retrieved by HRR2 does not compensate for its inability to identify relevant prior art. Similarly, the failure of ReQ-ReC, despite processing the same volume of documents as FullRecall, demonstrates its limitations in scenarios that require high recall. Table 9 indicates the rank of each target patent for each $P_{1\_UO}$, $P_{2\_UO}$, $P_{3\_UO}$, $P_{4\_UO}$ and $P_{5\_UO}$ within their respective ranked retrieved set from FullRecall.

**Table 9:** Ranking Positions of Examiner-Cited Target Patents for Patents under Observation: $P_{1\_UO}$, $P_{2\_UO}$, $P_{3\_UO}$, $P_{4\_UO}$ and $P_{5\_UO}$ within FullRecall's final ranked result set.

| Patent_under_observation | | Number of Target patents (examiner_citations) | Target patents (examiner_citations) | Rank |
|---|---|---|---|---|
| $P_{1_{UO}}$ | US2019053227A1 | 10 | EP3454474 A1 | 193 |
| | | | US20180262242 A1 | 1331 |
| | | | US20180324768 A1 | 1439 |
| | | | US20190014481 A1 | 788 |
| | | | US20190053227 A1 | 174 |
| | | | US20190363843 A1 | 1660 |
| | | | US20200037297 A1 | 694 |
| | | | US20200059867 A1 | 1319 |
| | | | US20220279537 A1 | 227 |
| | | | US20230098368 A1 | 1402 |
| $P_{2_{UO}}$ | US20200124681A1 | 4 | WO2017038387 A1 | 295 |
| | | | US20140302355 A1 | 895 |
| | | | EP3346542 A1 | 276 |
| | | | US20120176096 A1 | 70 |
| $P_{3_{UO}}$ | US10825607 B2 | 6 | US20130192860 A1 | 934 |
| | | | US20090192667 A1 | 348 |
| | | | US6072265 A | 28 |
| | | | US4945269 A | 967 |
| | | | US4631959 A | 34 |
| | | | US4075603 A | 1030 |
| $P_{4_{UO}}$ | US12168226 B2 | 5 | US20140066318 A1 | 1323 |
| | | | US20130116128 A1 | 1473 |
| | | | US20130053256 A1 | 926 |
| | | | US20120245053 A1 | 323 |
| | | | CA2473308 A1 | 1198 |
| $P_{5_{UO}}$ | US8976843 B2 | 7 | US20050063345 A1 | 207 |
| | | | US20040246998 A1 | 912 |
| | | | US8588326 B2 | 1273 |
| | | | US20040131007 A1 | 1568 |
| | | | US8102832 B2 | 1463 |
| | | | US20070097927 A1 | 149 |
| | | | US7379742 B2 | 148 |

Beyond ranking accuracy, the high recall performance of the proposed FullRecall framework is particularly important in practical patent search scenarios. In domains such as intellectual property, where missing even a single relevant document can lead to significant legal, financial, or strategic consequences, retrieval completeness is not merely beneficial; it is essential.

For inventors, the failure to identify prior art before filing a patent application can result in substantial wasted investment. Drafting, filing, and prosecuting a patent application involve considerable costs, including attorney fees, official filing charges, and responses to office actions. If the application is ultimately rejected due to overlooked prior art, these resources are lost. Additionally, such a rejection may delay product development timelines and expose the invention to competitive risk. The proposed FullRecall framework assists inventors in several keyways:

- Assessing Patentability – It enables accurate assessment of patentability by identifying existing prior art that may require modifications to the invention.
- Avoid Unnecessary Filing Costs – Patent applications involve significant expenses; a complete prior art search prevents wasted resources on inventions that may lack novelty.

- Strengthen Patent Claims – Knowing the prior art landscape allows inventors to draft patent a more robust claims and thus, minimizes rejection.
- Mitigate Infringement Risks – A thorough prior art analysis helps inventors navigate potential conflicts and avoid infringing on existing patents.

Inventors do not typically have predefined IPC codes for their inventions; Instead, they rely heavily on retrieval systems that can accurately capture semantically relevant documents. The proposed FullRecall framework, through its use of ranked noun phrases and semantic similarity, offers more reliable and comprehensive search capabilities compared to classification-based or keyword-only methods.

Patent examiners, whose role includes ensuring that patents are granted only to truly novel and inventive technologies, also can benefit significantly from the proposed FullRecall framework. A thorough prior art search is crucial to:

- Prevent the Issuance of Invalid Patents – Overlooking relevant prior art can lead to the granting of invalid patents, which can later be challenged and revoked.
- Supporting a rigorous and comprehensive patent examination – Examiners must identify all existing prior art to accurately assess whether a patent meets the required novelty and inventive step.
- Safeguard Against Potential Infringement Cases – Undiscovered conflicting patents may give rise to future legal disputes.

Patent litigation is often triggered by a prior art that was overlooked during the examination process; potentially resulting in costly reexaminations, invalidations, commercial delays and legal disputes. If invalidated, it not only imposes legal costs and commercial disruptions but also erodes trust in the ability of the patent office to uphold the standards of patentability. The proven ability of the proposed FullRecall framework to retrieve all relevant patents helps ensure that such oversights do not occur.

The comparative results between the proposed FullRecall framework and the baseline methods: HRR2 [3] and ReQ-ReC [2], clearly show that the proposed framework is more reliable when high recall is a necessity. HHR2 retrieves fewer documents but fails to capture a significant portion of relevant patent documents. On the other hand, whilst ReQ-ReC retrieves the same number of documents as the proposed framework, it often fails in capturing the relevant prior art. The proposed FullRecall framework consistently achieves 100% recall and ranks the relevant patents meaningfully within the result set; ensuring that no critical prior art is overlooked as well as enabling more efficient patent examination and decision making.

Given the importance of completeness in prior art retrieval, the proposed FullRecall framework addresses a critical need for both inventors and examiners; with its comprehensive coverage supporting strategic decision-making for inventors and strengthening the reliability of the examination process for patent offices. In applications where no relevant document can be missed, the proposed FullRecall framework offers a level of assurance that is essential.

## 5. Conclusion and Future Works

Effective patent prior art search depends not just on precision, but on the ability to retrieve all relevant information. This study highlights the value of recall-oriented retrieval methods that can uncover documents that traditional systems often miss. The improved recall demonstrated by this study presents a practical path forward for building more reliable, comprehensive search solutions in intellectual property and beyond.

The proposed FullRecall method leverages prior knowledge in the form of a patent's IPC codes description and applies natural language processing techniques to extract meaningful phrases. Then these phrases help to highlight the noun phrases from the query patent to narrow down the dataset. The retrieved set then undergoes semantic analysis and a ranking process, producing a refined list of documents with 100% recall. Overall, this study combines traditional keyword-based retrieval with more advanced semantic analysis using Bert for patents, coupled with a dedicated ranking scheme. This methodological blend of exact term matching with semantic understanding offers a robust framework for patent prior art search for both patent examiners and inventors. The experimental evaluation showed FullRecall outperformed the baseline methods HRR2 [1] and ReQ-ReC [2] in all test cases by achieving 100% recall. HRR2 [1] recall values across five test cases were 10%, 25%, 33.3%, 0%, and 14.29%, while ReQ-ReC [2] showed 50% for first test case, 25% for second test case, and 0% for third, fourth and fifth test case. These promising results demonstrate that the proposed FullRecall method can assist patent inventors and examiners to reliably achieve the desired recall, ensuring that no relevant prior art is overlooked, thereby strengthening the pre-filing patentability search and patent examination process and reducing potential legal and monetary risks.

In future work, we aim to automate the currently manual process of query formulation at the intermediary stage. This automation could significantly enhance efficiency and reduce the reliance on expert intervention. Another important direction is the inclusion of larger and more diverse patent datasets, which would enable a more thorough assessment of the robustness and adaptability of the proposed FullRecall approach across various domains. Additionally, we plan to extend our evaluation to full patent documents, rather than focusing solely on limited sections. This would allow for a more comprehensive and realistic evaluation of performance. However, such an extension would require careful consideration of the additional computational resources needed.